*Review Article*

# Dusty Blue Supergiants: News from High-Angular Resolution Observations

## W. J. de Wit,[1] R. D. Oudmaijer,[2] and J. S. Vink[3]

[1] *European Southern Observatory, Alonso de Cordova 3107, Casilla, Vitacura, 19001 Santiago, Chile*
[2] *School of Physics & Astronomy, University of Leeds, Woodhouse Lane, Leeds LS2 9JT, UK*
[3] *Armagh Observatory, College Hill, Armagh BT61 9DG, UK*

Correspondence should be addressed to W. J. de Wit; wdewit@eso.org

An overview is presented of the recent advances in understanding the B[e] phenomenon among blue supergiant stars in light of high-angular resolution observations and with an emphasis on the results obtained by means of long baseline optical stellar interferometry. The focus of the review is on the circumstellar material and evolutionary phase of B[e] supergiants, but recent results on dust production in regular blue supergiants are also highlighted.

## 1. Introduction

B-type stars of various luminosity classes and evolutionary phases can demonstrate the B[e] phenomenon. The stars are surrounded by material with a high equator-to-pole density contrast and a relatively large fraction of dust. The circumstellar material gives rise to line emission from permitted and forbidden transitions and excess radiation in the infrared. The flattened, axisymmetric geometry was unambiguously established by polarimetric observations that revealed scattering by both dust and electrons [1, 2].

The origin of the circumstellar material (CSM), its connection to the evolutionary phase, and the star's properties pose an intriguing problem. The CSM could originate from an evolved star through an, as yet, unidentified mass-loss process or it could constitute the accretion disk of a young star (e.g., [3]). A subclassification of B[e] stars was introduced based on the suspected evolutionary phase/luminosity class in order to compare like-with-like and facilitate the identification of the physical processes at play [4].

The B[e] phenomenon is particularly striking in a select group of massive blue supergiants to which we refer as sgB[e], following Lamers et al. [4]. These objects have bolometric luminosities between a few $10^4$ up to $10^6$ $L_\odot$. An exemplary and relatively homogeneous group is the Magellanic Cloud sgB[e] stars [5–8]. The galactic sgB[e] (and candidates) are listed in Kraus [9] but this group is necessarily less homogeneous because of the members' distance uncertainties.

It has long been recognized that the sgB[e]s may play an important role in understanding the evolution of (single) high-mass stars in terms of mass-loss and angular momentum evolution. Yet, massive star evolution is poorly understood because it is rapid while the observational characteristics of different evolutionary phases can be quite similar. For instance, the sgB[e]s have observational commonalities with the luminous blue variables (LBVs) and the yellow hypergiants. The uncertain nature of sgB[e]s goes hand-in-hand with the unknown nature of the parent group of the blue supergiants (e.g., [10]); the sgB[e] CSM can shed light on this particular issue.

The sgB[e]s show a two-component stellar wind. A fast (1000 km s$^{-1}$), hot (C IV, Si IV), low density, line-driven wind is present in the polar region. The fast wind is similar to that of normal B supergiants. The expansion velocity (100 km s$^{-1}$) and temperature are much lower in the equatorial region and, assuming constant mass flux, this implies a much denser wind. The historical interpretation involves a fast rotating supergiant with a rotationally induced, latitudinal stratified wind by virtue of the bistability and the Von Zeipel effects [5, 11, 12]. In order for this scenario to be effective the star has to rotate at >50% of break-up velocity. Rotational velocities have been measured for four sgB[e]s to date, two of which rotate at >50% of critical velocity (see [7, 8, 13, 14]).

The maturing of long baseline optical interferometry has offered the opportunity to probe the CSM of sgB[e]s at high spatial and spectral dynamic ranges. This has brought important new insights into the nature of these objects. A review of the topic with a greater emphasis on the technical background is provided by Millour [15] and for lower mass evolved objects by Chesneau [16]. In this contribution we present a review of recent work in understanding the B[e] phenomenon among supergiants using high-angular resolution, mostly spectrointerferometric observations. A brief overview is also provided of relevant developments in the field of the blue supergiants.

## 2. The Blue Supergiants

We briefly discuss three blue supergiant (BSG) properties that are or could be relevant for the sgB[e]: evolutionary phase indicators, rotational velocities, and evidence for dust production.

*2.1. Evolutionary Phase.* The OBA supergiants are the evolutionary descendants of the high-mass O-type dwarfs. They are expected to enter this phase directly following the main sequence or after a red supergiant (RSG) phase on a blue loop if the star has sufficiently large initial mass [17]. The limiting mass for a blue loop trajectory depends on the mixing and mass-loss processes. As these two processes are not well constrained, major uncertainties exist whether the BSGs are in a pre- or post-RSG phase. Saio et al. [18] introduce a method that allows differentiating between the two possibilities based on the star's pulsational properties. According to this approach, the BSG subgroup of the $\alpha$ Cyg variables is supposedly in a post-RSG phase. However, the observed CNO abundance cannot be made in agreement with the model prediction, although this could be reconciled by the treatment of convection in the models [19]. Also the absence of IR excess does not speak in favour of a post-RSG phase either for the $\alpha$ Cyg variables, indeed none of the BSGs of the Crowther et al. [20] sample display strong IR excesses, which is exemplified in Figure 1. The figure shows the mid-IR colours for BSGs and sgB[e]s as measured by the *WISE* satellite [21]. We have indicated the expected colours for various dust temperatures and bound-free and free-free (bf-ff) excesses. Nonetheless, because such late-B and A-type supergiants display spectroscopic variability (timescales of 10 to 50 days) related to structures at the base of the wind (e.g., [22]), they could generate a complex environment especially when interaction with a slow RSG wind takes place [23]. The H$\alpha$ variability at 1 milliarcsecond spatial resolution as reached by the *CHARA* interferometric array provides direct spatial information of the inhomogeneous and unsteady environment of the $\alpha$ Cyg variables [24].

*2.2. Rotation.* The rotational velocity distribution of the blue supergiant population is stepwise in $\log(g) - v\sin(i)$ space:

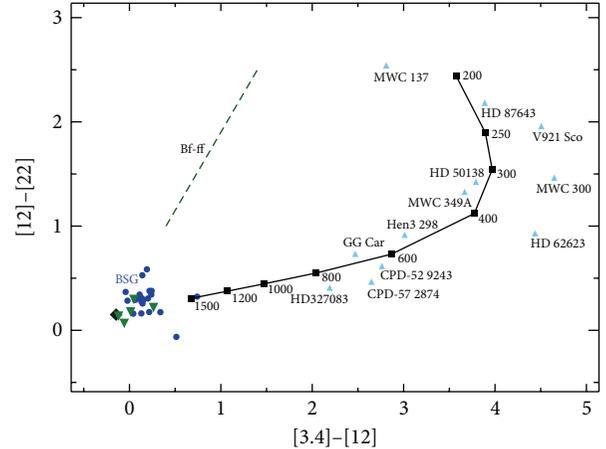

Figure 1: WISE mid-IR colour-colour diagram for galactic BSGs. The full lines are the colours produced by the simple addition of two blackbodies representing the star (of 20 000 K) and the dust (with a variety of temperatures as indicated) for a constant $L_{IR}/L_{bol}$ of 0.6%. The sgB[e] candidates (blue triangles [9]) and regular BSGs (blue dots) are indicated from the sample studied by Crowther et al. [20]. The colours for stars with excess bf-ff emission were determined empirically. Green triangles are $\alpha$ Cyg variables, *namely*, HD 64760, HD 163899, HD 46769, HD 152249, and HD 199478.

very few fast rotators (>100 km$^{-1}$ s$^{-1}$) are found with gravities less than $\log(g) \approx 3.2$. This functional form could be the result of the end of main sequence evolution for BSGs [25]. Alternatively, the sudden decrease in $v\sin(i)$ could be caused by bistability (BS) braking, that is, the increase in mass-loss at the BS-jump resulting in an accelerated angular momentum loss [10]. The latter scenario would see the removal of a significant fraction of the star's angular momentum by mass-loss already on the main sequence. If this is a dominant effect, then it may pose problems for the fast rotation scenario as the cause for sgB[e]s if rotation is the same for BSGs and sgB[e]s.

*2.3. Dust.* A small number of galactic BSGs are known to have equatorial rings with diameters of a few tenths of parsec, *namely*, Sher 25 in the massive young cluster NGC 3603 [26] and the seemingly isolated blue supergiant SBW 1 [27]. Both nebulae contain dust and are remarkable in their similarities to SN1987A [28, 29]. A spherical RSG wind which interacts with an hour-glass BSG wind would reproduce grossly the observed morphology of the rings and nebulae [30]. However, the observed N/O abundance and considerations regarding the stellar mass are incompatible with the idea that the stars were RSGs [27, 31, 32]. Moreover, dusty RSG winds are rather nonspherical and clumpy as shown by high-angular resolution in the near-IR (≤100 mas) and mid-IR ($\leq 1''$, [33–35]). An explanation for the few examples of BSG equatorial ring nebulae in terms of a wind-wind interaction encounters some observational difficulties. On the other hand, the alternative scenario for explaining the nebulae by binary interactions has gained some observational support from the complex equatorial environment of the eclipsing blue supergiant binary RY Scuti [36].

The properties and effects observed in BSGs can be relevant for understanding the processes at play among the sgB[e]s. Conversely, the sgB[e]s have also a story to tell that bears on normal BSGs by virtue of the properties of the circumstellar material.

## 3. Recent Advances in Observed Properties of sgB[e] Stars

We list observational characteristics of sgB[e] with an emphasis on the results obtained in the last 8 years, since the meeting dedicated to the B[e] phenomenon [37]. A particular emphasis is given to results with optical interferometers like the *VLTI* [38] and *CHARA* [39]. Our aim is to remain strictly with supergiants and avoid other evolutionary phases that show the B[e] phenomenon.

*3.1. The Dusty Material.* The production of dust in the CSM and its continuous presence close to the star are one of the intriguing properties of sgB[e]s. The dust has a typical temperature of 1000 K (i.e., lower than sublimation [7, 8]) with some instances of warm (200–300 K) dust ([40], see Figure 1).

*Spitzer Space Telescope (SST)* IR spectra of the well-defined sample of Magellanic Cloud sgB[e] stars are presented in Kastner et al. [41]. Their results show that several objects show crystalline silicate dust features reminiscent of the features seen in the disks of premain sequence stars. The production of crystalline dust requires physical conditions which are met in objects with circumstellar or circumbinary disks and in cool objects with strong mass-loss, like AGB and RSG stars [42]. For the sgB[e]s, the presence of crystalline dust argues in favour of rather long-lived or quasipermanent presence of the dust particles within the circumstellar structure as opposed to temporary, continuous outflowing material [41]. Yet, the fact that the CSM is characterized by mass-loss processes (and not by accretion) is corroborated by the large dust shells seen on linear scales of tens of parsec also detected by the *SST* and by the parsec scale polar nebulae and filamentary shells in the light of H$\alpha$ [43]. In comparison to other massive star evolutionary phases, it is worth mentioning that some LBVs (bar $\eta$ Car) also show crystalline silicates (e.g., [42]). Nonetheless the distinct near-IR colours between LBVs and sgB[e]s (see Figure 13 in [44]) probably demonstrate that the geometry of the emitting structure is different. Similarly, LBVs and sgB[e]s show a clear difference in hydrogen recombination line ratios which may suggest different geometries as well [45].

Neither a two-component wind nor the Keplerian disk scenario for the CSM can properly reproduce the infrared excess emission of the sgB[e] [46]. The author calculated the bf-ff and the dust continuum emission assuming a critical density for dust grain growth. Given the radial temperature structures of the CSM, the author finds that the dust emission (as compared to that of the LMC sgB[e] R 126) could not be reproduced unless a significant alteration of the radial density stratification is imposed. Mediating this problem, Kraus et al. [47] argued in favour of a largely hydrogen neutral sgB[e] CSM practically all the way down to the star. In this way ample surface area could be dust producing and remedy the underestimated dust emission as calculated by Porter [46]. Yet, numerical calculations by Carciofi et al. [48] (see also [49]) find that even in the densest part of the equatorial region the temperature does not fall below the dust condensation temperature of 1500 K. As the angular scales involved are on milliarcsecond scales, interferometric observations can settle the issue of the exact radius of the dust sublimation.

The galactic object CPD-57 2874 was the first sgB[e] for which the CSM was directly probed at milliarcsecond scale using the VLTI [50]. The dust continuum geometry agreed well with that determined by polarimetric measurements lending further support to higher dust densities in the equatorial region. Radiative transfer modelling was employed in modelling a larger body of interferometric data by the same authors finding that the 10 $\mu$m visibilities indicate a relatively "large" inner dust edge of radius ~10 AU. The radius of the inner edge constitutes tens of stellar radii [51]. A comparable scale for the dust inner edge was estimated recently by Cidale et al. [52] based on *VLTI/MIDI* N-band data for the sgB[e] CPD-52 9243. Finally, using the 3-telescope beam-combiner *VLTI/AMBER*, Wang et al. [53] probed the continuum emission in the near-IR. Supplying a working model derived from modelling spectroscopic data [54], the authors derive an inner radius for the hot dust of ~30 AU.

The first cases of direct size measurements of the dust emitting region near sgB[e]s indicate a dust free zone within orbits of roughly tens of stellar radii. This is consistent with the absence of hot dust (e.g., Figure 1) and suggests a geometry similar to a dust ring rather than a dust disk. Further evidence that is consistent with the idea of a dust ring rather than a disk is provided by the IR continuum modelling in [55, 56].

*3.2. The Gaseous Material.* The dense CSM provides shielding against the stellar radiation field such that molecules and dust can form. Emission from the rovibrational $\Delta \nu =$ 2–0 transitions of the carbon monoxide molecule has been detected in about 50% of the sgB[e]s [57–59]. The emission originates on AU-scales in the CSM and is an excellent probe for its dynamics [60]. Attaining a spatial resolution of 3 mas, Wheelwright et al. [61] employ the spectrointerferometric capabilities of *VLTI/AMBER* to investigate the CO bandhead emission from the sgB[e] HD 327083. These data are supplied with near-simultaneous high resolution spectra ($R = 50\,000$) provided by *VLT/CRIRES* resolving the bandhead in velocity space. The velocity field imprint is that of material in a Keplerian disk orbiting at a radius of ~3 AU (2 mas). Consequently, given the spatial resolution of the interferometer, the emitting region remained spatially unresolved. Nonetheless, any relative photocentre displacement between continuum and line emission of tens of microarcsecond (depending on projected baseline length) will show up as a displacement in the fringe phase (see [62]). This is the signal shown in Figure 2. It indicates a 0.15 mas displacement over the bandhead emission with a precision of 30 $\mu$as. The agreement between the Keplerian model and observations in the spatial and spectral domains is compelling evidence that the CO bandhead emission of HD 327083 originates in a Keplerian disk. In contrast, the model of an equatorial outflow cannot reproduce the differential phase signal well [61].

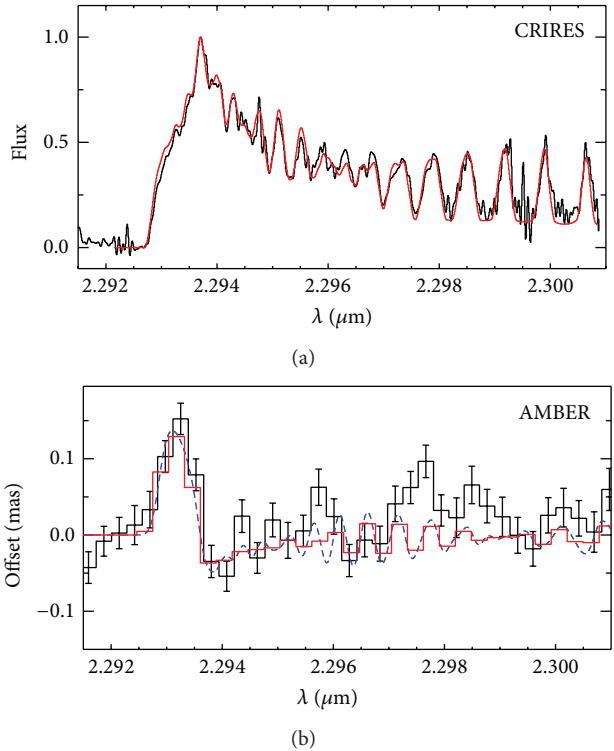

FIGURE 2: *VLTI/AMBER* and *VLT/CRIRES* observations of CO bandhead emission of HD 327083 [61]. (a): Observed flux spectrum (black) and model fit (red) based on a Keplerian disk. (b): Differential phase spectrum (histogram) with the Keplerian model prediction at a spectral resolution of *VLT/CRIRES* (blue) and of *VLTI/AMBER* (red). Reproduced with permission from Astronomy & Astrophysics, © ESO.

By spatially resolving both the Br$\gamma$ emission and the dust continuum with *VLTI/AMBER*, Millour et al. [63] demonstrate that the ionized gas located at ~2 AU is in Keplerian rotation whereas the dust inner rim is found at ~6 AU in the A[e] supergiant HD 62623. Their reconstructed synthesis images show an inner cavity in the ionized gas distribution. Similar conclusions for the kinematics are drawn indirectly by means of spectroscopy for other sgB[e]s. Cidale et al. [52] find Keplerian rotation for the detached cold CO ring of CPD-52 9243. Liermann et al. [64] conclude that the CO material for two LMC sgB[e]s cannot extend down to the stellar surface, because of the observed temperature of the CO gas. Quasi-Keplerian rotation in a ring is also the conclusion reached for LHA 115-S 65 by Kraus et al. [14]. On the basis of high-density tracers, Aret et al. [65] find that the line profiles of both [O I] and [Ca II] indicate that the discs or rings of high-density material in sgB[e]s are in Keplerian rotation.

The dedicated studies iterate that the gas is located in a Keplerian rotating ring interior to the dust location. The CO diagnostic provides also insight into the temporal stability of the disk. Zickgraf et al. [6] mention that the Magellanic Cloud sgB[e] shows in general no marked photometric variability, but the galactic sgB[e] group seems to be more variable albeit less homogeneous [66]. Wheelwright et al. [61] find that the CO bandhead was stable over a period of 5 months in HD 327083 stretching the temporal baseline of the observations. Thureau et al. [67] report the stability of the resolved ring-like dust structure over a 7-year period around the luminous HMXB[e] CI Cam using the *IOTA* interferometer. It is worth noticing that this period follows immediately the pronounced outburst of the source in 1998. The stability of the dominant structure is accompanied by significant, low amplitude (few tenths) photometric variability longward of the *V* band and corresponds to flux levels on average somewhat brighter than the preoutburst emission [68]. In general, appropriate light curve coverage of sgB[e]s has been sparse, but one good example is LHA 115-S 18. Over a period of 16 years the object displays strong photometric variability in the optical [69]. Nonetheless, the low excitation emission lines are largely unaffected by this optical variability, possibly indicating that the equatorial "cool" material is not the agent for this variability [69]. Moreover, the IR spectrum in Kastner et al. [41] suggests the presence of crystalline dust in this particular source S 18 which would also support the longevity of the equatorial structure.

The strongest evidence so far for variability in disk/ring properties is the sudden (<9 months) appearance of CO bandhead emission in LHA 115-S 65, reported by Oksala et al. [70]. This appearance is enigmatic and difficult to understand. If the equatorial outflow is driven by viscosity and material is launched in Keplerian orbit by fast rotation combined with some mass-ejection process (e.g., nonradial pulsations and small scale magnetic fields; see [71]), then classical Be type continuum variability can be expected among the sgB[e]s. The zone responsible for the strongest variability in bf-ff emission in a gaseous disk is located in the optical thick part close to the stellar surface with typical colour variability related to transition between the fractions of optical thick and thin material [72, 73]. The absence of any reports of Be star variability in these stars can thus be taken as a further indication that the sgB[e] "disk" is a ring *or* that the bf-ff emission is only a small fraction of the total excess emission dominated by the dust. Interferometric phase variation across the Br$\gamma$ line indicates that the photons in this transition are dominated by the wind in early-type sgB[e] [50] and the profile is quite distinct from the spatial imprint of a rotating Keplerian disk as typically seen in classical Be stars with optical interferometry (e.g., [74]) and spectroastrometry [75].

In general, sgB[e] variability time-scales and amplitudes are not well-known. If the equatorial outflow is produced and governed by similar physics as the ones operating in classical Be stars, then the presence of the dust and polar wind complicates the picture substantially. By the same token, binarity will not be a governing factor in the process of creating a sgB[e] star [76].

## 4. Binarity and Circumbinary Rings

The evidence for a correlation between the occurrence of the B[e] phenomenon and binarity in B-type supergiants is growing, although a causal relation has not been established. By means of high-angular resolution observations or otherwise, confirmation of *circumbinary* Keplerian rings as the most probable geometry for the equatorial outflows has been provided for four objects, *namely*, MWC 300, HD 327083, HD 62623, and GG Car [53, 63, 77, 78]. Two sgB[e]s with similar circumstellar physics have not revealed a binary nature, at least so far, *namely*, CPD-52 9243 and CPD-57 2874 [51, 52].

The question is whether BSGs have winds that are strongly latitudinally dependent when rotating fast or if the phenomenon of sgB[e] is purely induced by binarity. In case of the former it may be caused by, for example, bistability as mentioned earlier. In case of the latter processes like nonconservative mass transfer (e.g., [56]) or the gravitationally focusing of the wind in the equatorial plane (or both) are responsible for retaining material in orbit around the system. We briefly expand on two objects that could play an important role.

(1) The primary in HD 62623 is a common A2 supergiant rotating at 0.3 to 0.6 times the critical velocity. As detailed in Meilland et al. [79], these values are not high enough to create a sufficient pole-to-equator density contrast according to Pelupessy et al. [11], although a second BS jump at $T_{\text{eff}} \approx 10\,000$ K [80] is not investigated in that work. Until the reality of a second BS jump is verified, the production of a dusty circumbinary disk would otherwise be hard to understand without the aid of a companion object [81].

(2) GG Car is an eccentric binary with a circumbinary disk. The eccentric orbit [82] will be maintained (and not circularized) when enhanced mass-loss is induced during close periastron passages [83]. The material will be captured in the equatorial plane as a circumbinary disk by the gravitational pull of the secondary (e.g. [84]). GG Car is not the only dusty eccentric system with a circumbinary disk. What could be the closest analogon to the CSM in sgB[e] are the dusty circumbinary disks of post-AGB stars. These objects show the presence of crystalline silicates and CO emission lines indicative of temporally stable circumbinary disks [85]. The presence of dust in these systems is always related to binarity [86] and the binary orbits are generally quite eccentric. The disk/ring may provide a mean to pump eccentricity (see [87]), but other mechanisms have been proposed to explain the eccentricity, like the enhanced mass-loss during periastron passage mentioned before [83].

Inspired by the presence of a circumbinary disk and dust composition analogies between sgB[e] and post-AGB, it has been suggested that sgB[e]s are objects in a post-RSG evolutionary phase. However this suggestion can be discarded given the recent results on the sgB[e] $^{13}$C abundance. Kraus [9] shows that the evolutionary phase can be strongly constrained by means of the $^{12}$C/$^{13}$C isotopic ratio. In sgB[e] this ratio can be probed using the $^{12}$CO and $^{13}$CO bandheads in the near-IR. Based on this approach, Oksala et al. [44] conclude that the sgB[e]s are likely in a pre-RSG phase of their evolution. A pre-RSG phase for the sgB[e] merits consideration because of the examples of dusty rings in regular BSG (Section 2) and the similarities to the massive blue supergiant binary RY Scuti. In these BSG systems the abundance and system orbits are inconsistent with a post-RSG phase. The properties of objects thought to be in a post-RSG phase are discussed in Oudmaijer et al. [88]. One of these objects is the famous yellow hypergiant IRC +10420 and, contrary to the AU-scale environment of the sgA[e] HD62623, it shows ionized emission which is best reconciled with an hour-glass wind geometry [89] rather than a rotating disk as in HD 62623 [63]. These findings tend to favour a pre-RSG phase for the BSGs and sgB[e]s, at least for those objects on evolutionary tracks that cross the HR-diagram to the RSG region, that is, the ones with luminosities less than approximately $\log(L_{\text{bol}}/L_\odot) = 5.8$.

## 5. Summary and Final Remarks

The new high-angular resolution data provide important observational input into the discussion whether the sgB[e] phenomenon is intrinsically related to the blue supergiant mass-loss and its importance for massive star evolution in general or whether the circumstellar environment is shaped by binary interaction. To increase the number of genuine sgB[e]s and verify the nature of the many unclassified B[e] stars, high-angular resolution observations continue to play an important role (e.g., [48, 90, 91]). The case of the well-known source V921 Sco which appears in the list of galactic sgB[e] candidates is illustrative. Here, spatially and spectrally resolved observation of the CSM leads to uncovering properties which are strongly in favour of a premain sequence nature for V921 Sco by means of determining the stellar mass and dynamics [92].

We can summarize the findings for the sgB[e] which have been studied in detail as follows. The kinematics of the equatorial material is revealed to be in Keplerian rotation rather than in radial expansion for a significant fraction of the sgB[e]s. There is a significant incidence of short-period binaries among the sgB[e]s and the dynamical and spatial information indicate that the geometry of the equatorial material is a circumbinary disk, with much lower densities of material at smaller radii. The presence of crystalline silicates and temporal monitoring indicate that the CSM in the equatorial region seems to be rather stable in time. The isotopic ratio of $^{12}$C/$^{13}$C indicates a pre-RSG phase for most of the sgB[e]s.

Not only increasing the number of sgB[e]s studied at high-angular resolution but also determining the incidence of short-period binaries in normal BSG will help in determining the cause for this interesting phenomenon. The ultimate aim is to reveal whether the binary companion is a sufficient reason for inducing a long-lasting B[e] phenomenon in a small subset of supergiant B-type stars or whether rotation needs to play a role.

## Conflict of Interests

The authors declare that there is no conflict of interests regarding the publication of this paper.

## Acknowledgments

This publication makes use of data products from the Widefield Infrared Survey Explorer, which is a joint project of the University of California, Los Angeles, and the Jet Propulsion Laboratory/California Institute of Technology, funded by the National Aeronautics and Space Administration. This research has made use of NASA's Astrophysics Data System.

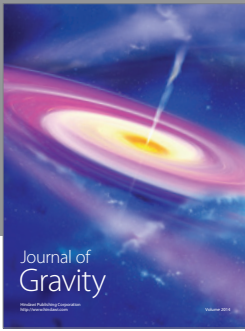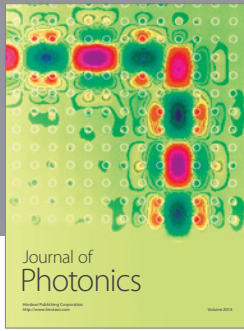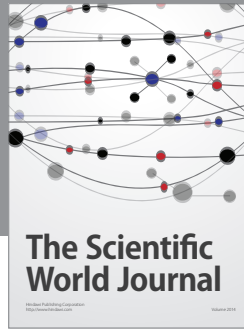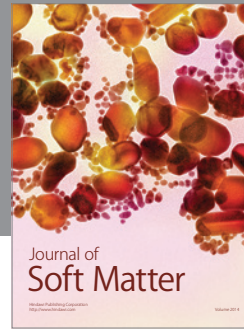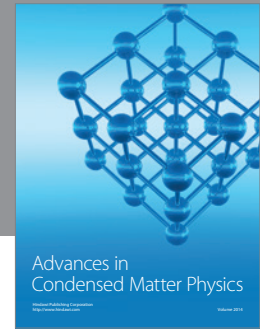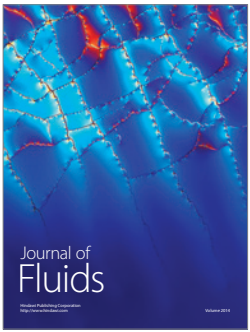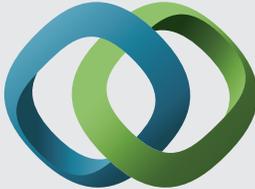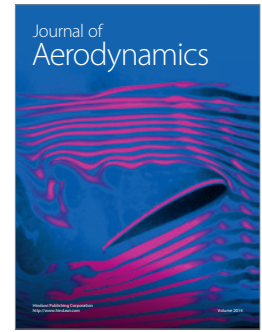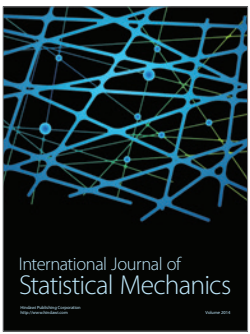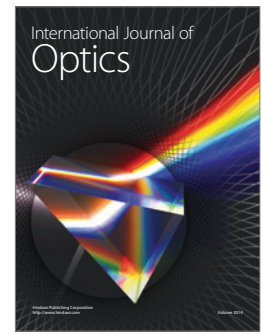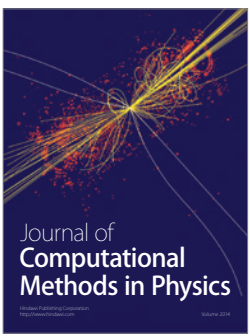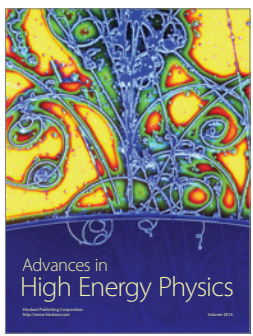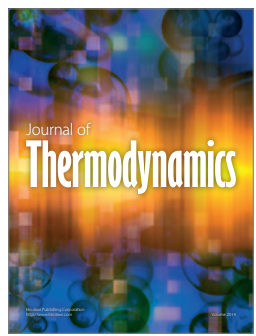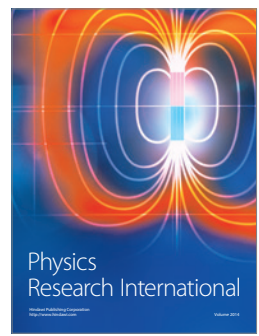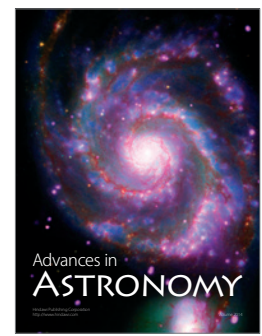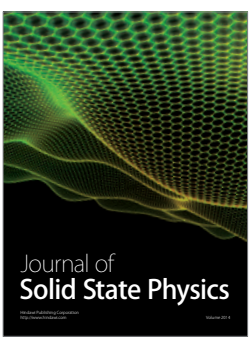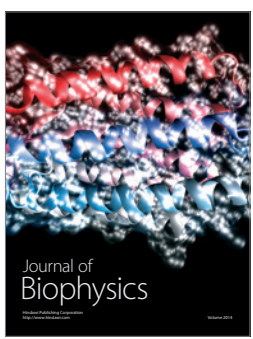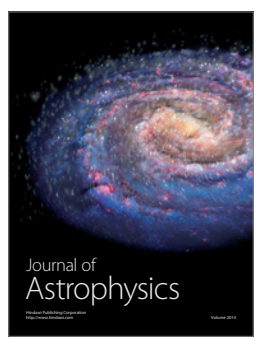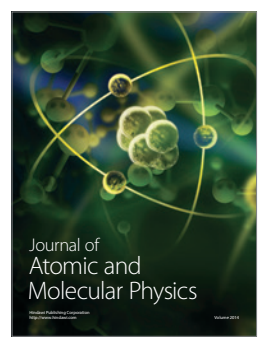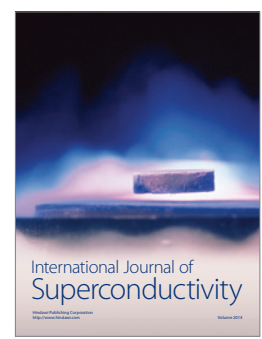